\documentclass[
twocolumn,
]{ceurart}

\usepackage{amsmath,amsfonts,amssymb}
\usepackage{epigraph}

\usepackage[normalem]{ulem}
\useunder{\uline}{\ul}{}

\usepackage{listings}
\begin{document}

%%
%% Rights management information.
%% CC-BY is default license.
\copyrightyear{2022}
\copyrightclause{Copyright for this paper by its authors.
  Use permitted under Creative Commons License Attribution 4.0
  International (CC BY 4.0).}

%%
%% This command is for the conference information
\conference{EvalRS at CIKM 2022}

%%
%% The "title" command
\title{Triplet Losses-based Matrix Factorization for Robust Recommendations}

%%
%% The "author" command and its associated commands are used to define
%% the authors and their affiliations.
\author[1]{Flavio Giobergia}[%
orcid=0000-0001-8806-7979,
email=flavio.giobergia@polito.it
]
% \cormark[1]
\address[1]{Department of Control and Computer Engineering, Politecnico di Torino, Turin, Italy}

% \author[3]{Ilaria Tiddi}[%
% orcid=0000-0001-7116-9338,
% email=i.tiddi@vu.nl,
% url=https://kmitd.github.io/ilaria/,
% ]
% \fnmark[1]
% \address[3]{Vrije Universiteit Amsterdam, De Boelelaan 1105, 1081 HV Amsterdam, The Netherlands}

% \author[4]{Manfred Jeusfeld}[%
% orcid=0000-0002-9421-8566,
% email=Manfred.Jeusfeld@acm.org,
% url=http://conceptbase.sourceforge.net/mjf/,
% ]
% \fnmark[1]
% \address[4]{University of Skövde, Högskolevägen 1, 541 28 Skövde, Sweden}

%% Footnotes
% \cortext[1]{Corresponding author.}

%%
%% The abstract is a short summary of the work to be presented in the
%% article.
\begin{abstract}
Much like other learning-based models, recommender systems can be affected by biases in the training data. While typical evaluation metrics (e.g. hit rate) are not concerned with them, some categories of final users are heavily affected by these biases. In this work, we propose using multiple triplet losses terms to extract meaningful and robust representations of users and items. We empirically evaluate the soundness of such representations through several ``bias-aware'' evaluation metrics, as well as in terms of stability to changes in the training set and agreement of the predictions variance w.r.t. that of each user. 
\end{abstract}

%%
%% Keywords. The author(s) should pick words that accurately describe
%% the work being presented. Separate the keywords with commas.
\begin{keywords}
  recommender systems \sep
  matrix factorization \sep
  contrastive learning
\end{keywords}

%%
%% This command processes the author and affiliation and title
%% information and builds the first part of the formatted document.
\maketitle

\section{Introduction}
Recommender systems are a fundamental part of almost any experience of online users. The possibility of recommending options tailored to each individual user is one of the key contributors to the success of many companies and services. The metrics that are commonly used in literature to evaluate these models (e.g. hit rate) are typically only concerned with the overall quality of the model, regardless of the behaviors of such models on particular partitions of data. This results in recommender systems typically learning the preferences of the ``majority''. This in turn implies a poorer quality of recommendations for users/items that belong to the long tail of the distribution.
In an effort to steer the research focus to addressing this problem, the EvalRS challenge \cite{evalrs}. This challenge, based on the RecList framework \cite{chia2022beyond}, proposes a recommendation problem with a multi-faceted evaluation, where the quality of any solution is not only evaluated in terms of overall performance, but also based on the results obtained on various partitions of users and items. 
In this paper, we present a possible recommender system that addresses the problem proposed by EvalRS. The solution is based on matrix factorization by framing an objective function that aligns users and items in the same embedding space. The matrices are learned by minimizing a loss function that includes multiple triplet losses terms. Differently from what is typically done (i.e. aligning an anchor user to a positive and a negative item), in this work we propose additionally using triplet terms for users and items separately.

The full extent of the challenge is described in detail in \cite{evalrs}. In short, the goal of the challenge is to recommend songs to a set of users, given their previous interactions with other songs. The provided dataset is based on a subset of the openly available LFM-1b dataset \cite{schedl2016lfm}. The source code for the proposed solution has been made available on GitHub \footnote{\url{https://github.com/fgiobergia/CIKM-evalRS-2022}}.

% Information about both users (e.g. age, gender, country of origin, behaviors) is available, as well information on the songs (authors, title, albums where they have been published). 

% The rest of the paper is organized as follows. Section \ref{sec:method} presents the proposed solution, which is then evaluated in Section \ref{sec:results}. Section \ref{sec:fails} reports some of the approaches that have not produced improvements in the model but that have been deemed promising. Finally, Section \ref{sec:conclusions} draws some conclusions on the challenge and the results achieved.

\section{Methodology}
\label{sec:method}
In this section we present the proposed methodology, highlighting the main aspects of interest. No data preprocessing has been applied to the original data, although some approaches have been attempted (see Section \ref{sec:fails}). 
The proposed methodology, as explained below, allows ranking all items based on estimated compatibility with any given user. We produce the final list of $k$ recommendations by stochastically selecting items from the ordered list of songs, weighting each song with the inverse of its position in the list.

\subsection{Loss definition}
Matrix factorization techniques have long been known to achieve high performance in various recommendation challenges \cite{koren2009matrix}. This approach consists in aligning vector representations for two separate entities, users and items (songs, in this case). This alignment task is a recurring one: a commonly adopted approach to solving this problem is through the optimization of a triplet loss \cite{schroff2015facenet}. 

A triplet loss is a loss that requires identifying an anchor point, as well as a positive and a negative point, i.e. points that should either lie close to (positive) or far from (negative) the anchor point. 

Users and songs can thus be projected to a common embedding space in a way that users are placed close to songs they like and away from songs they do not like. This can be done by choosing a user as the anchor, and two songs as the positive and negative points. A reasonable choice for the positive song is one that has been listened by the user. The choice for the negative song is trickier. Random songs, or songs not listened by the user are possible choices. However, more sophisticated strategies can be adopted to choose negative points that are difficult for the model to separate from the anchor. These are called hard negatives and have been shown in literature to be beneficial to the training of models \cite{xuan2020hard}. 

We decided to use a simple policy for the selection of a negative song: a negative song for user $u$  is extracted from the pool of songs that have been listened by one of the nearest neighbors of $u$ and have not been listened by $u$. By doing so, we aim to reduce the extent to which the model relies on other users' preferences to make a recommendation. The concept of neighboring users is obtained by comparing the similarity between embedding representations of all users. Due to the computational cost of this operation, it is only performed at the beginning of each training epoch. 

We can thus define the triplets $(u^a_i, s^p_i, s^n_i)$ to be used for the definition of a triplet loss. Here, $u^a_i$ is used to represent the vector for the anchor user, whereas $s^p_i$ and $s^n_i$ represent the vectors for the positive and negative songs respectively.

% All users and songs representations are part of the $U$ and $S$ matrices that are being learned.

Similar approaches where users are aligned to songs they did or did not like are Bayesian Personalized Ranking (BPR) \cite{rendle2012bpr}, where negative songs are sampled randomly, and WARP \cite{weston2010large}, where negative items are sampled so as to be ``hard'' (based on their proximity of the anchor w.r.t. the positive item). To improve the robustness of the representations built, we are additionally interested in aligning similar songs and similar users. To this end, we introduce two additional triplet terms to the loss function, one that is based on $(s^a_i, s^p_i, s^n_i)$ and one on $(u^a_i, u^p_i, u^n_i)$. Based on the previously defined concepts, we choose $s^a_i$ as a song listened by $u^a_i$, and $u^p_i$ and $u^n_i$ as users who respectively listened to $s^p_i$ and $s^n_i$. Other alternatives have been considered, but were ultimately not selected due to a higher computational cost. 

We define the final loss as:
\begin{equation}
\label{eq:loss}
\begin{split}
\mathcal{L} = \sum_i w_i (max\{d(u^a_i, s^p_i) - d(u^a_i, s^n_i) + m_0, 0\} + \\  \lambda_1 max\{d(u^a_i, u^p_i) - d(u^a_i, u^n_i) + m_1, 0\} + \\ \lambda_2 max\{d(s^a_i, s^p_i) - d(s^a_i, s^n_i) + m_2, 0\})
\end{split}
\end{equation}

Where $d(\cdot)$ is a distance function between any pair of vectors. In this work, the cosine distance is used. $m_j$ is a margin enforced between positive and negative pairs. In this work, since all elements are projected to a common embedding, we used $m_0 = m_1 = m_2$. Finally, $w_i$ is a weight that is assigned to each entry, which is discussed in Subsection \ref{ssec:weight}.

\begin{figure}
    \centering
    \includegraphics[width=.8\linewidth]{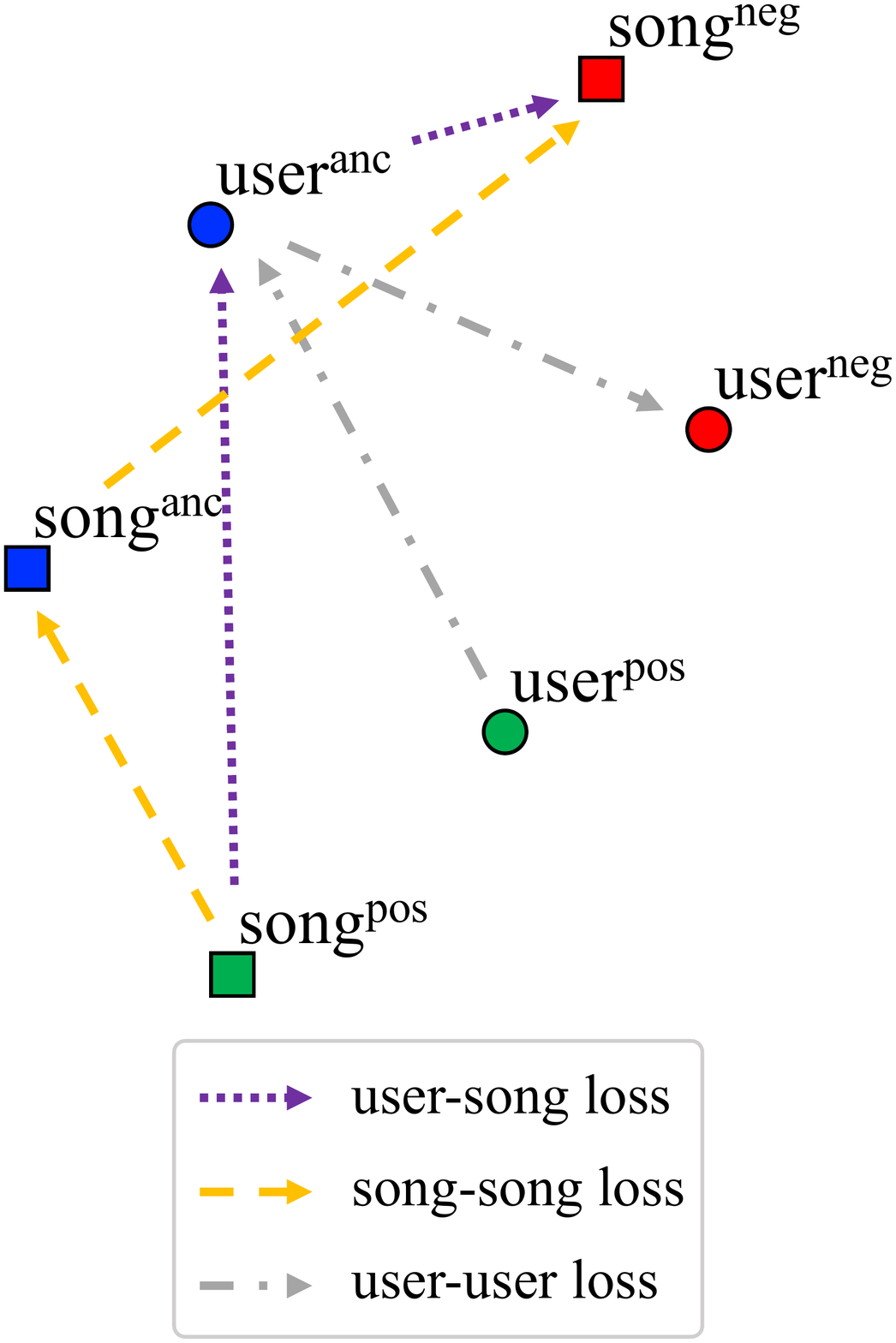}
    \caption{Representation of the action of each part of the loss on the vectors. Arrow directions represent whether elements are pulled towards or pushed away from the anchors.}
    \label{fig:loss}
\end{figure}

Figure \ref{fig:loss} summarizes the effect of the various terms of the loss on the embedding vectors learned.

\subsection{Popularity weight}
\label{ssec:weight}
To make the minority entities more relevant, we adopted a weighting scheme that modulates the previously described loss so as to weigh rows more if they belong to ``rarer'' entities and less for common ones. In accordance with \cite{evalrs}, we identified five factors to be kept into account. Based on these, a coefficient has been defined for each entry in the training set. The final weight is given by a weighted average of these coefficients. The following is a list of factors, along with the way the respective coefficients have been computed (logarithms are used for factors that follow a power law distribution). All coefficients are normalized to sum to 1 across the respective population.
\begin{itemize}
    \item \textit{Gender} ($\theta_{gender}$): in accordance with the original dataset, a relevance coefficient is provided for the categories  male, female, and undisclosed \footnote{This simplified perspective on gender does not reflect that of the author}. The coefficient is proportional to the inverse of the occurrences of each gender in the list of known users.
    \item \textit{Country} ($\theta_{country}$): the coefficient related to the country is calculated as the inverse of the logarithm of the number of occurrences of the specific country of the users in the training set.
    \item \textit{Artist popularity}  ($\theta_{artist}$): a proxy for the popularity of an artist is obtained by the number of times songs by that artist have been played in the training set. The inverse logarithm of this quantity is used as coefficients.
    \item \textit{Song popularity}  ($\theta_{song}$): a proxy for the popularity of a song is provided by the number of times that song has been played in the training set. The inverse logarithm of this quantity is used as coefficients.
    \item \textit{User activity}  ($\theta_{activity}$): the overall activity of a user can be quantified in terms of the number of songs that they have listened to across the training set. The inverse logarithm of this quantity is used as coefficients.
\end{itemize}

The weighted sum of the above-mentioned coefficients constitutes the weight $w_i$ in Equation \ref{eq:loss}. The weights used for each coefficient have been searched as a part of the tuning of the model and are presented in Section \ref{sec:results}.

% Figure \ref{fig:weights} shows the distribution of these weights according to the best performing configuration found. 

% \begin{figure}
%     \centering
%     \includegraphics[width=\linewidth]{images/weights.pdf}
%     \caption{Distribution of weights for all training points in a randomly generated fold of the dataset. Larger values are assigned to a small set of points that are represented the least in terms of the identified factors.}
%     \label{fig:weights}
% \end{figure}

\subsection{Model initialization}
The initial values assigned to the users' and items' vectors greatly affects the entire learning process. A good initialization can make the convergence process faster and/or allows reaching a better minimum. We used initial vectors for users and items based on an adaptation of the word2vec algorithm \cite{mikolov2013distributed}. We built a corpus of sentences, one for each song known, composed of users who listened to that song, artists and albums, all in the form token-type=token-value (e.g. song=1234). We then trained word2vec to learn representations for all of the tokens involved. We used as initial vectors the vectors obtained for the users and songs tokens.

Word2vec places tokens close in the embedding space based on their adoption in similar contexts. For this reason, based on the definition of sentences, this approach already brings close users with similar tastes -- in terms of songs, artists, albums, as well as similar songs -- in terms of users that listened to them, artists the produced them, albums they are found in. 

\subsection{Model consistency}
\label{ssec:consistency}
As we will discuss in Section \ref{sec:results}, we empirically observed that the proposed solution presents high variance in the performance obtained across the various folds. While this is not directly measured as a part of the core metrics of EvalRS, we still believe it is important to account for this aspect in a well-rounded evaluation. 

We therefore introduce the \textit{consistency} metric, which quantifies the variance of the model when tested across multiple folds, or datasets. A higher variance in performance would be associated with a lower consistency (or higher inconsistency). For a single metric, the consistency could be defined as the variance of the metric across the folds. However, when multiple metrics are involved (as is the case with this competition), a normalization step should be introduced. We thus instead use the coefficient of variation, defined as the standard deviation divided by the mean value, to quantify the inconsistency of a model with respect to a metric $m$. We compute the consistency for a metric as 1 - inconsistency. The overall consistency is therefore computed as the mean consistency across all metrics:

\begin{equation}
    c = \frac{1}{|M|} \sum_{m \in M} \left(1 - \frac{\sigma_m}{|\mu_m|} \right)
\end{equation}

Where $M$ represents the set of all metrics used, while $\mu_m$ and $\sigma_m$ are the arithmetic mean and standard deviation computed over all the folds, for a metric $m$. We use the absolute value of the mean to make the results comparable regardless of sign. Alternatively, the ratio $\sigma_m^2/\mu_m^2$ could be used to assign a lower penalty in case of small deviations. The maximum possible efficiency, 1, would be assigned to a model that presents the same exact performance across all folds for all metrics. Section \ref{sec:results} reports the consistency, measured in these terms, for the proposed solution. 

\subsection{Variance agreement}
Different users may have different interests in terms of variety. In the ``music'' context a user may, for example, listen to songs from very few authors, whereas others may be more interested in a wider variety of artists. A similar concept may be applied to other contexts (e.g. in terms of brand loyalty for products). It is therefore desirable that a recommender system should provide a wider variety of recommendations for users that are inclined to them, and vice versa. We introduce the concept of \textit{variance agreement} w.r.t. a variable, which quantifies how the variance in recommendations correlates to each user's interest in variance, as dictated by their previous interactions, in terms of the variable of interest. In this context, we use the artists that produced songs as the variable of interest.

We quantify the variance of a set of songs as the Gini impurity over that set, where each song is mapped to the respective artist. We can thus assign an impurity to any given user, $g_u$, as the impurity within the set of songs they listened to in the training set. For that same user, we can define the model's impurity, $g_{\hat{u}}$, as the impurity of the set of $k$ songs recommended by the model for that user.

If $g_u$ is low, the user listens to a limited set of artists (if 0, the user has only listened to one artist in all of its interactions). Similarly, if $g_{\hat{u}}$ is low, the model is recommending songs from a limited set of artists. 

To measure the agreement between users and model's variance, we compute the Pearson correlation on the paired data $[ (g_u, g_{\hat{u}} )\;|\;u \in \mathcal{U} ]$, with $\mathcal{U}$ being the set of all users. 

Note that the variance agreement is not concerned with the correctness of the predictions (e.g. whether the artists the user listens to are the same ones being recommended) -- that information may be quantified by other metrics concerned with the accuracy of models, rather than their suitability over a heterogeneous set of user.

\section{Experimental results}
\label{sec:results}
In this section we present the results obtained in terms of the main metrics identified by \cite{evalrs}, as well as some additional considerations on the proposed solution.

The model has been trained and fine-tuned to identify well-performing values for the main hyperparameters. The best configuration of parameters found is reported in Table \ref{tab:hyper}.

% The training has been performed on batches of 512 elements, for a total of 2 epochs. The dataset is only processed twice before the model reaches peak performance -- any further training has not been observed to produce any improvement. We attribute this behavior to the reduced complexity of the proposed model (i.e. two matrices, without any non-linearity being introduced). We instead observed a wide range of behaviors when tuning the parameters $\lambda_1$ and $\lambda_2$.

\begin{table}[h]
\centering
\begin{tabular}{cc}
\hline
\textbf{Parameter} & \textbf{Value} \\ \hline
$d$ & 128 \\ \hline
$\lambda_1$ & 2.5 \\
$\lambda_2$ & 2.5 \\
$m_0 = m_1 = m_2$ & 0.25 \\ \hline
$\theta_{gender}$ & 5 \\
$\theta_{country}$ & 100 \\
$\theta_{artist}$ & $10^4$ \\
$\theta_{song}$ & $10^5$ \\
$\theta_{activity}$ & $10^4$ \\ \hline
\end{tabular}
\caption{List of main hyperparameters configured. Horizontal lines separate categories of hyperparameters that have been tuned together.}
\label{tab:hyper}
\end{table}

Table \ref{tab:results} presents the results obtained on the challenge leaderboard for the top 10 entries.

% Please add the following required packages to your document preamble:
% \usepackage[normalem]{ulem}
% \useunder{\uline}{\ul}{}
\begin{table*}[]
\centering
\resizebox{\linewidth}{!}{
\begin{tabular}{ccccccccccc}
\hline
\multicolumn{1}{c}{\textbf{Solution}} & \multicolumn{1}{c}{\textbf{Score}} & \multicolumn{1}{c}{\textbf{Hit rate}} & \multicolumn{1}{c}{\textbf{MRR}} & \multicolumn{1}{c}{\textbf{\begin{tabular}[c]{@{}c@{}}Country\\ (MRED)\end{tabular}}} & \multicolumn{1}{c}{\textbf{\begin{tabular}[c]{@{}c@{}}User activity\\ (MRED)\end{tabular}}} & \multicolumn{1}{c}{\textbf{\begin{tabular}[c]{@{}c@{}}Track popularity\\ (MRED)\end{tabular}}} & \multicolumn{1}{c}{\textbf{\begin{tabular}[c]{@{}c@{}}Artist popularity\\ (MRED)\end{tabular}}} & \multicolumn{1}{c}{\textbf{\begin{tabular}[c]{@{}c@{}}Gender\\ (MRED)\end{tabular}}} & \multicolumn{1}{c}{\textbf{\begin{tabular}[c]{@{}c@{}}Being\\ less wrong\end{tabular}}} & \multicolumn{1}{c}{\textbf{\begin{tabular}[c]{@{}c@{}}Latent\\ diversity\end{tabular}}} \\ \hline
team\#1 & {\ul \textbf{1.702570}} & 0.015484 & {\ul \textbf{0.005859}} & -0.004070 & -0.006932 & {\ul \textbf{-0.002044}} & -0.001688 & -0.000956 & {\ul \textbf{0.424817}} & {\ul \textbf{-0.121655}} \\
team\#2 & {\ul 1.552977} & {\ul 0.016065} & {\ul 0.001727} & {\ul \textbf{-0.003727}} & {\ul \textbf{-0.002913}} & {\ul -0.002307} & -0.001047 & {\ul \textbf{-0.000692}} & {\ul 0.363927} & -0.296403 \\
Proposed solution & 1.330388 & 0.015565 & 0.001677 & -0.004036 & -0.003504 & -0.004444 & \textbf{-0.000867} & -0.000797 & 0.281863 & -0.272944 \\
team\#4 & 1.184669 & {\ul 0.021619} & {\ul 0.002044} & -0.005366 & -0.004417 & {\ul -0.003191} & -0.001542 & -0.001299 & {\ul 0.320594} & -0.317348 \\
team\#5 & 1.138580 & {\ul 0.018819} & 0.001071 & -0.005213 & -0.005174 & -0.005043 & -0.001234 & -0.002774 & 0.280727 & {\ul -0.244437} \\
team\#6 & 1.028222 & 0.015006 & 0.001127 & -0.005448 & -0.007534 & -0.005261 & -0.001202 & -0.003369 & {\ul 0.316226} & -0.309870 \\
team\#7 & 0.752576 & {\ul 0.017874} & 0.001655 & -0.006193 & -0.010749 & -0.004483 & -0.002132 & -0.004541 & {\ul 0.322594} & -0.324841 \\
team\#8 & 0.429596 & {\ul 0.018173} & 0.001632 & -0.005482 & -0.011190 & -0.007588 & -0.002513 & -0.004329 & {\ul 0.322767} & -0.324794 \\
team\#9 & -1.154413 & {\ul 0.032359} & {\ul 0.002054} & -0.010624 & -0.013020 & -0.021992 & -0.003122 & -0.008458 & {\ul 0.295921} & -0.330054 \\
baseline & -1.212213 & {\ul \textbf{0.036343}} & {\ul 0.003694} & -0.009016 & -0.022362 & -0.011082 & -0.007150 & -0.006084 & {\ul 0.375774} & -0.307977
\end{tabular}
}
\caption{Performance in terms of the various metrics adopted in \cite{evalrs} for the top 10 participants (MRR = Mean Reciprocal Rank, MRED = Miss Rate Equality Difference). Underlined are the solutions that outperform the proposed one in term of the specific metric. In bold are the best performers for each metric.}
\label{tab:results}
\end{table*}

The table clearly shows that the proposed solution does not outperform the top achievers in most aspects. It is unfortunately impossible to currently compare the techniques adopted by the rest of the participants to draw more meaningful conclusions. However, from the results, it is clear that the proposed solution fares reasonably well in the partition-based metrics, while it struggles to achieve comparable performance in terms of other metrics (most notably, the ``being less wrong'' one). 

It is also interesting to note how other solutions still have performance trade-offs, since there is no clear solution that outperforms all others across all metrics. This is due to the multi-faceted nature of the overall score function. 

Despite the efforts made toward reducing the effect of the dataset imbalances on the final model, we still observed that the performance of the model is not always consistent. In other words, there is a relatively high variance in the performance across the various folds. 

Table \ref{tab:cons} shows the overall consistency of the model, as well as the consistency computed for each metric. This provides a very interesting perspective on the strengths and weaknesses of the proposed solution. In particular, the model is highly inconsistent for some of the fairness-oriented metrics -- as highlighted by the low consistency obtained for track popularity and gender. While this does not necessarily imply poor performance, it is a symptom that the model may be susceptible to fluctuations in performance as the dataset used for training is changed. Other metrics, such as the behavioral and the ``standard'' ones, show instead a more consistent behavior.

\begin{table}[]
\centering
\begin{tabular}{lc}
\hline
\textbf{Metric} & \textbf{Consistency} \\ \hline
Hit rate & 0.9579 \\
MRR & 0.8885 \\
Country (MRED) & 0.8057 \\
User activity (MRED) & 0.9312 \\
Track popularity (MRED) & 0.6938 \\
Artist popularity (MRED) & 0.7828 \\
Gender (MRED) & 0.4573 \\
Being less wrong & 0.9949 \\
Latent diversity & 0.9976 \\ \hline
Overall & 0.8344 \\ \hline
\end{tabular}
\caption{Consistency obtained for each metric, as measured across $k=4$ folds.}
\label{tab:cons}
\end{table}

We additionally evaluated the proposed methodology in terms of variance agreement for the ``artist'' variable. We achieved an agreement of 0.2479, whereas a random model would achieve $\approx 0$. This indicates that the model does take into account, to some extent, the individual user's variance preference. However, there is room for improvements in these terms.

\subsection{Ablation study}
To understand the effect of the various choices made, we introduce an ablation study where we remove some portions of the proposed methodology. In particular, we study the situations where (1) no user-user loss is consider, (2) no item-item is considered, (3) a random initialization is used instead of word2vec and (4) all training records are weighted the same, regardless of their rarity. Table \ref{tab:ablation} reports the results obtained, in terms of the final score, for all situations. From this we can observe that all proposed approaches bring a benefit to the overall result, with the removal of the additional loss terms being the most important. It should be noted, however, that this ablation study has been carried out using the hyperparameters that produced the best performance for the ``full'' approach. As such, the ablated performance may be affected by a lack of hyperparameters fine-tuning, thus possibly resulting in a lower score.

\begin{table}[]
\centering
\begin{tabular}{lc}
\hline
\textbf{Ablated term} & \textbf{Score} \\ \hline
No user-user loss & -100\footnotemark \\
No item-item loss &  0.5292 \\
No word2vec initialization &  0.8554 \\
No weighting scheme & 0.8427 \\
Full approach & \textbf{1.3304} \\
\hline
\end{tabular}
\caption{Ablation study of the proposed solution. Performance is reported in terms of the overall score adopted for the competition.}
\label{tab:ablation}
\end{table}

\section{Failed approaches}
\label{sec:fails}
\epigraph{I have not failed. I've just found 10,000 ways that won't work.}{Thomas A. Edison}

In this section we describe some attempts that have been made, but that have not brought any improvement in terms of performance. 
% \begin{itemize}

\textit{Entity resolution}: the list of known songs contains some duplicates. We tried using a naive entity resolution approach (songs with matching artists and matching titles are considered to be the same song). Since this problem affected only a small fraction (a few percent) of songs, the ER step did not produce any significant improvement and has thus been discarded.

\textit{Dataset resampling}: we attempted to resample the training set with a weighting scheme similar to the one already used to weigh each training sample based on their uniqueness. Worse performance have been observed as a result of this approach: it can be argued that the reason for this is that the resampling outright removes some samples from the training set (points that are never sampled), whereas using a weight for each row makes sure that all rows are seen during training. 

\textit{Data augmentation}: to increase the breadth of the data available, we tried to synthesize new user-song interactions, to be then used for training. In particular, we first quantified the proclivity of users to listen to a limited number of artists, by means of the Gini impurity (the more homogeneous the choice of artists, the lower the Gini index). We can then sample users based on this factor, and add user-song relationships, where songs are chosen to belong to the most ``likely artists'' (i.e. the artists that are more commonly listened by each sampled user).
% \end{itemize}
\footnotetext[3]{A score of -100 has been assigned to solutions that did not reach a hit rate of 0.015.}

\section{Conclusions}
\label{sec:conclusions}
In this paper we presented a possible solution to the EvalRS challenge. The solution uses matrix factorization based on multiple triplet losses combined together to align users and songs in the same space. A weighting scheme has been introduced to assign more importance to uncommon users/items -- thus improving the quality of the model in terms of fairness. By introducing the consistency metric, we show some of the main weaknesses of the proposed approach: namely, the fact that it is not consistent w.r.t. some metrics. We consider this to be one of the main problems to be addressed.  We additionally covered some of the failed attempts made, in the hope that others will either not make the same mistakes, or figure out how to improve upon them.

%%
%% The acknowledgments section is defined using the "acknowledgments" environment
%% (and NOT an unnumbered section). This ensures the proper
%% identification of the section in the article metadata, and the
%% consistent spelling of the heading.
\begin{acknowledgments}
This work has been supported by the DataBase and Data Mining Group and the SmartData center at Politecnico di Torino.
\end{acknowledgments}

%%
%% Define the bibliography file to be used
\bibliography{sample-ceur}

\end{document}